\documentclass{article}
\usepackage[utf8]{inputenc}
\usepackage{graphicx}
\graphicspath{ {./images/} }
\usepackage{float}
\usepackage{hyperref}
\usepackage{rotating}
\restylefloat{table}
\usepackage{color}
\usepackage{enumitem}
\usepackage{subcaption}

\usepackage{graphicx}

\usepackage{geometry}
\title{A unified machine learning approach to time series forecasting applied to demand at emergency departments}
\author{M.A.C. Vollmer, B. Glampson, T. A. Mellan, S. Mishra, L. Mercuri, \\ C. Costello, R. Klaber, 
G. Cooke, S. Flaxman\footnote{Correspondence: s.flaxman@imperial.ac.uk}, S. Bhatt \\ ~ \\ 
MRC Centre for Global Infectious Disease Analysis, Imperial College London \\
Imperial College Healthcare NHS Trust \\
Faculty of Medicine, Imperial College London \\ 
Department of Infectious Disease, Imperial College London \\
Department of Mathematics, Imperial College London}
\date{June 2020}

\begin{document}

\maketitle

\begin{abstract}
There were 25.6 million attendances at Emergency Departments (EDs) in England in 2019 corresponding to an increase of $12$ million attendances over the past ten years \cite{NHS20, NHS17-18, NHS08-09}. The steadily rising demand at EDs creates a constant challenge to provide adequate quality of care while maintaining standards and productivity. Managing hospital demand effectively requires an adequate knowledge of the future rate of admission. Using $8$ years of electronic admissions data from two major acute care hospitals in London, we develop a novel ensemble methodology that combines the outcomes of the best performing time series and machine learning approaches in order to make highly accurate forecasts of demand, $1$, $3$ and $7$ days in the future. Both hospitals face an average daily demand of $208$ and $106$ attendances respectively and experience considerable volatility around this mean. However, our approach is able to predict attendances at these emergency departments one day in advance up to a mean absolute error of $\pm 14$ and $\pm 10$ patients corresponding to a mean absolute percentage error of $6.8$\% and $8.6$\% respectively. Our analysis compares machine learning algorithms to more traditional linear models. We find that linear models often outperform machine learning methods and that the quality of our predictions for any of the forecasting horizons of $1$, $3$ or $7$ days are comparable as measured in MAE. In addition to comparing and combining state-of-the-art forecasting methods to predict hospital demand, we consider two different hyperparameter tuning methods, enabling a faster deployment of our models without compromising performance. We believe our framework can readily be used to forecast a wide range of policy relevant indicators.

\end{abstract}

\section{Background}
In 2019 there were $25.6$ million attendances at emergency departments (EDs) in the UK, corresponding to $70,231$ patients attending every day \cite{NHS20}. The National Health Service (NHS) trusts across England are under very high pressure to maintain current standards and quality of care \cite{2019NHSKeyStatistics}. In fact, the rate of attendance has grown by $4.8$\% since 2018 and by $10.6$\% over the last $5$ years meaning that it is increasing at a rate faster than population growth thus putting high pressure on our health care system. Failure to make provisions for surges in demand can lead to overcrowding, which in turn has been linked to multiple adverse patient outcomes such as unfavourable patient satisfaction, poor quality of care and diseconomies of scale \cite{mccarthy2011overcrowding}. In order to decrease overcrowding, the NHS introduced a new operational standard in 2010, commonly-known as the ``four hour target'', requiring that at least $95$\% of patients attending EDs should either be discharged, admitted or transferred within $4$ hours of arrival. However, this target has not been met since 2014 and failure rates have reached a new high in January 2020 with $18$\% of patients at EDs waiting longer than four hours despite the fact that the overall number of attendances was lower in January 2020 than in January 2019. In fact $15.3$\% of patients spent more than $4$ hours in EDs in 2019 compared to $11.9$\% last year and $5.5$\% five years ago, making 2019 the year with the worst annual performance on record \cite{NHS20}. 
A shortage of staff is not the predominant cause for long waiting times and low quality of care \cite{silvester2004reducing}, rather, it is that the correct \textit{type} of staffing is not matched with patient demand. This inefficiency in staffing can have substantial impacts such as the $86,264$ of elective surgeries that had to be cancelled for non-clinical reasons on the day the patient was due to arrive in 2019. These cancellations leave NHS hospital trusts with lost costs of surgeons, anaesthetists and nurses as well as surgical session time and theatre capacity. Moreover, in the same year, the percentage of patients who had not been treated within $28$ days of cancellation decreased from $9.8$\% in 2018 to $8.8$\% in 2019, however, still failing one of the NHS's improvement objectives. In addition the NHS England advises that a $85$\% bed occupancy rate is the maximum safe level of occupancy and it advises that trusts should try and keep bed occupancy below $92$\%. However, $126$ out of $170$ trusts recorded bed occupancy above $85$\%, $58$ trusts rates above $92$\% in the third quarter of 2019, eight trusts had occupancy above $98$\% and one trust recorded $100$\% bed occupancy (NHS SitRep). 

Overall the NHS in England has spent around £$130$ billion in 2018/19 on the delivery of health services \cite{NHScosts}. A major fraction, $44.9$\% in the financial year 2016-2017 \cite{NHSfinances2018}, of this spending is due to the $1.2$ million people employed by NHS hospital and community health services. Between February 2018 and 2019 the number of doctors alone rose by $2.5$\% and by $10$\% during the past five years \cite{NHS-Staff}. This emphasises the fact that long waiting times may not simply result from a shortage of staff. Nonetheless, every year high agency and staffing costs are required to cover for staffing shortages. A key step to addressing the issue of staffing is the prediction of the rates of admission and the duration of stay at EDs (collectively referred to as ``demand''). In particular, an adequate prediction of demand and a better understanding of the reasons for demand is fundamental to the delivery of high quality care. 

\subsection{Previous work}
Despite its importance, methods for predicting rates of admission and understanding underlying dynamics have not been studied extensively in the literature. Existing methods have been limited to the application of classical time series forecasting methods. In an Australian study, Boyle et al \cite{boyle2012predicting} predict monthly, daily and four hourly demand at $27$ EDs in Queensland. Taking public holidays into account as predictors, the authors used an autoregressive integrated moving average (ARIMA) model as well as regression and exponential smoothing methods to  predict demand up to a mean absolute percentage error (MAPE) of $11$\% for daily admissions. McCarthy et al \cite{mccarthy2008challenge} predicted hourly presentations to American emergency departments while including several temporal, weather and patient factors on the number of hourly arrivals in order to characterise behaviour at EDs. Jones et al \cite{jones2008forecasting} used seasonal autoregressive integrated moving average (SARIMA), time series regression, exponential smoothing and artificial neural network models to forecast demand. The authors had access to two years of data and made predictions ranging from $1$ to $30$ days in advance for three hospitals in the USA. The authors found seasonal and weekly patterns to be most important for an accurate prediction and obtained a Mean Absolute Predictive Error (MAPE) of $9-13$\% depending on the facility. Champion et al \cite{champion2007forecasting} performed an analysis of monthly demand at an emergency department in regional Victoria. They applied exponential smoothing and Box-Jenkins methods to five years worth of admissions data. Hoot et al \cite{hoot2008forecasting} carried out a discrete event simulation in order to obtain forecasts for $2$, $4$, $6$ and $8$ hours into the future. The authors analysed waiting times, length of stay and bed occupancy.

\subsection{Our contribution}
We develop a novel, predictive framework to understand the temporal dynamics of hospital demand and we apply an exhaustive statistical analysis to daily presentations at EDs at St Mary's and Charing Cross Hospitals, evaluating a range of standard time series and machine learning approaches and ultimately developing our own unique approach. In contrast to existing studies, we do not only focus on the application of time series algorithms in order to characterise demand but develop a generic procedure that allows us to compare and combine both time series and machine learning algorithms in order to obtain an informative, more appropriate and consistently accurate approach to the prediction of demand. Our models have the ability to be retrained regularly and efficiently and are therefore a powerful tool for online platforms and near real time prediction. Using novel data from electronic logging systems from eight years of daily presentations to EDs at St Mary's and Charing Cross Hospitals in London, we construct a model that predicts the number of daily arrivals to both hospitals. Our analysis accounts for seasonal fluctuations, daily observed weather data and specific, pre-planned events indicated by staff at both EDs such as the yearly Notting Hill Carnival. Using our procedure can help the hospitals with the provision of the right staffing numbers and deploying resources in the most effective way. 

\section{Data sources}
\begin{figure}[H]
    \centering
    \includegraphics[scale=0.5,width=0.75\linewidth]{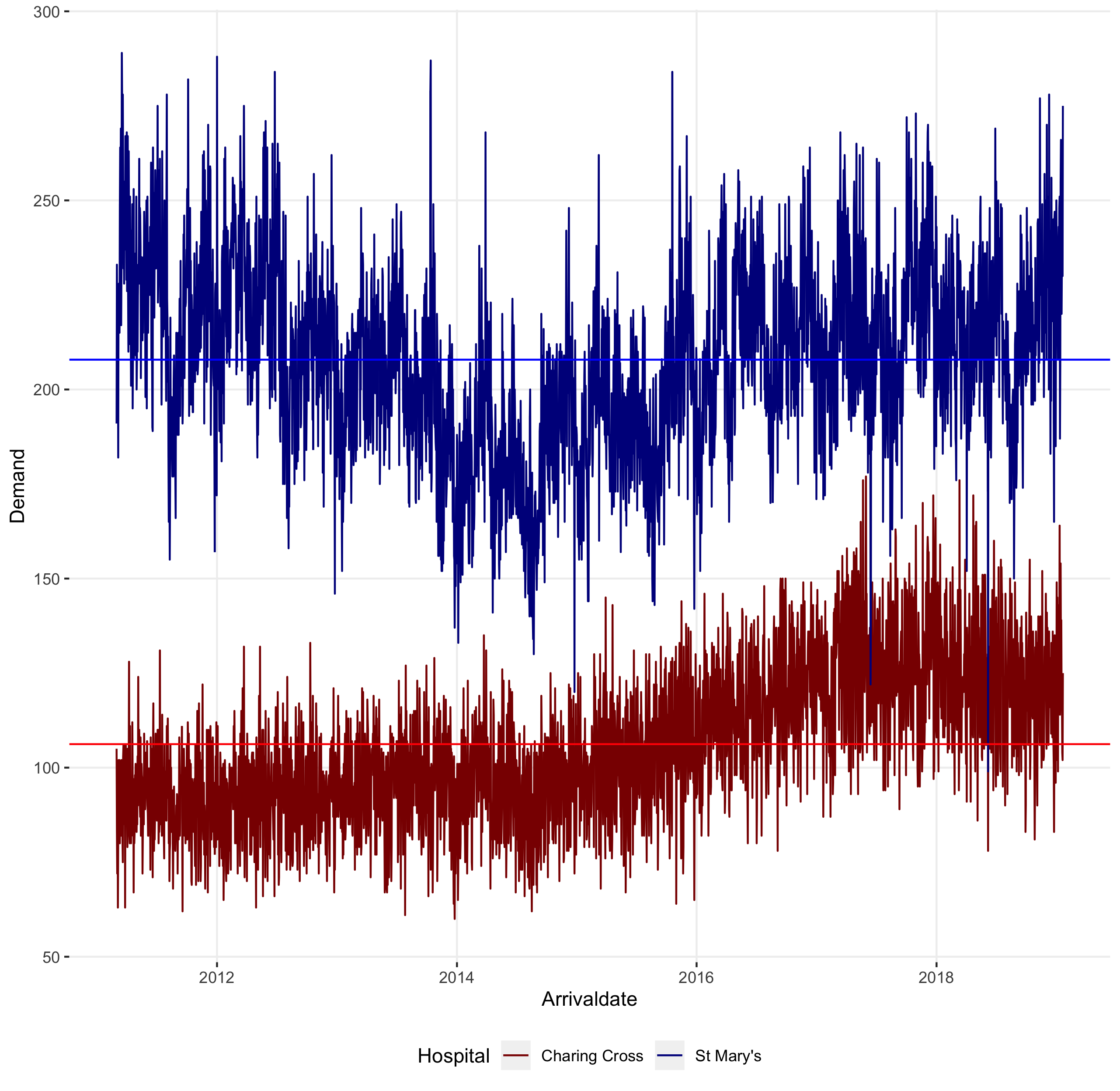}    
    \caption{Attendances at EDs per day for 2011 to 2019. The horizontal lines are the average demands.}
    \label{fig:timeseries}
\end{figure}

The St Mary's and Charing Cross Hospitals are part of the Imperial College Healthcare NHS Trust, one of $228$ NHS hospital trusts in England. St Mary's Hospital is the major acute care hospital for North West London housing a major trauma centre. Its ED has faced an average demand of $208$ patients (with a maximum of $289$ and a minimum of $99$) every day since 2011.  Charing Cross Hospital includes the serious injuries centre for West London as well as a hyper acute stroke unit. On average there have been $106$ (with a maximum of $177$ and a minimum of $60$) daily attendances at the ED since 2011. For our analysis, we had access to electronic data records of the number of daily attendances at the EDs for both hospitals from 2011 to 2018 (see Figure \ref{fig:timeseries}). In order to investigate the demand dynamics, we also collected data on school \cite{schoolholidays} and bank holidays \cite{bankholidays}, as well as on the weather and Google search volume for the word ``flu'' \cite{GoogleTrends-adjust,GoogleTrends} (see the Appendix for more details). Finally, experienced staff at the EDs of both hospitals provided us with a list of specific \textit{known} events in the locality that cause surge in demand (e.g. the Notting Hill carnival - an annual festival taking place in the catchment area of the hospital).  

\section{Exploratory data analysis}

\begin{figure}
  \begin{subfigure}[b]{0.45\textwidth}
    \includegraphics[width=\textwidth]{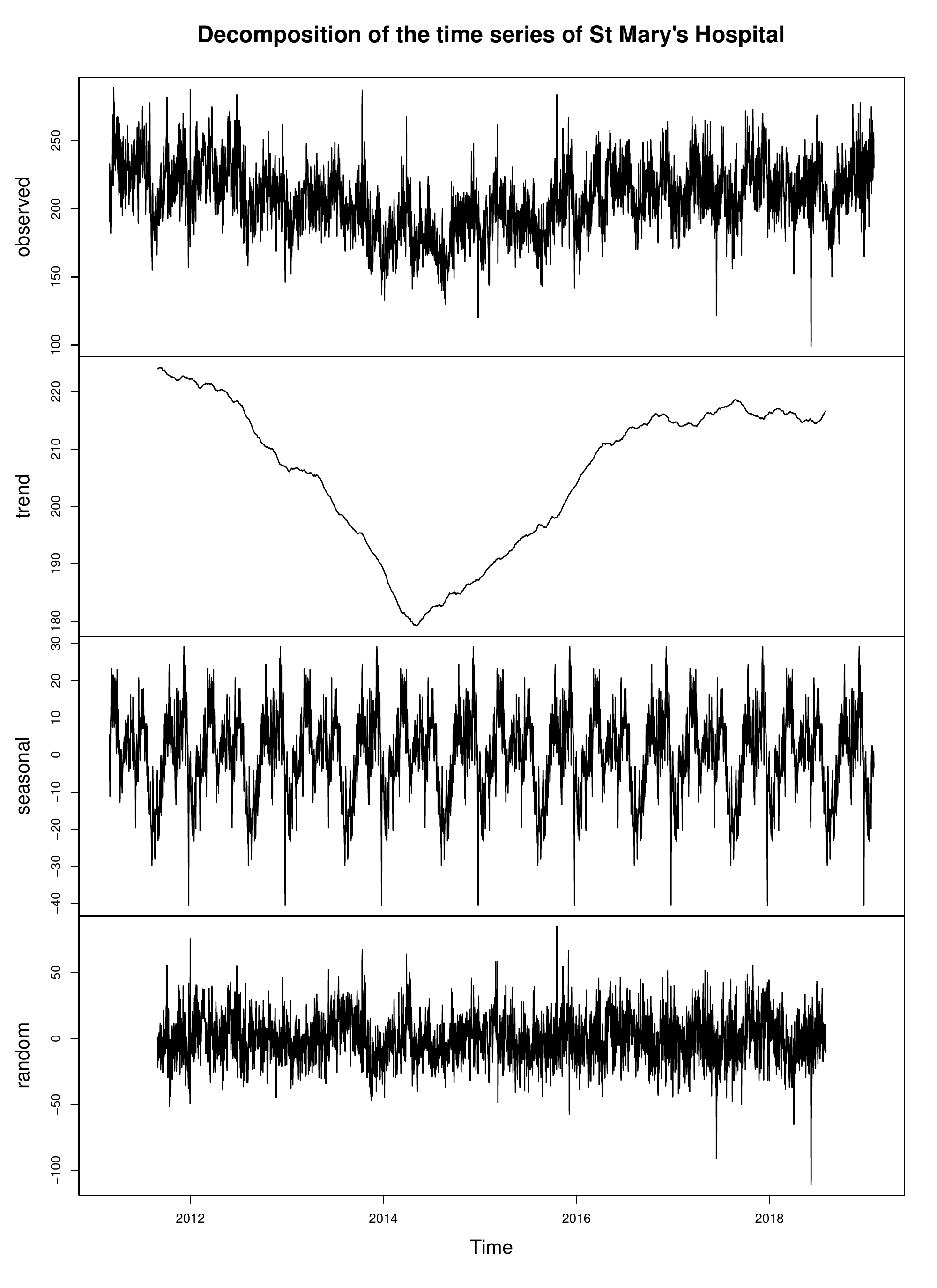}
  \end{subfigure}
  \begin{subfigure}[b]{0.45\textwidth}
    \includegraphics[width=\textwidth]{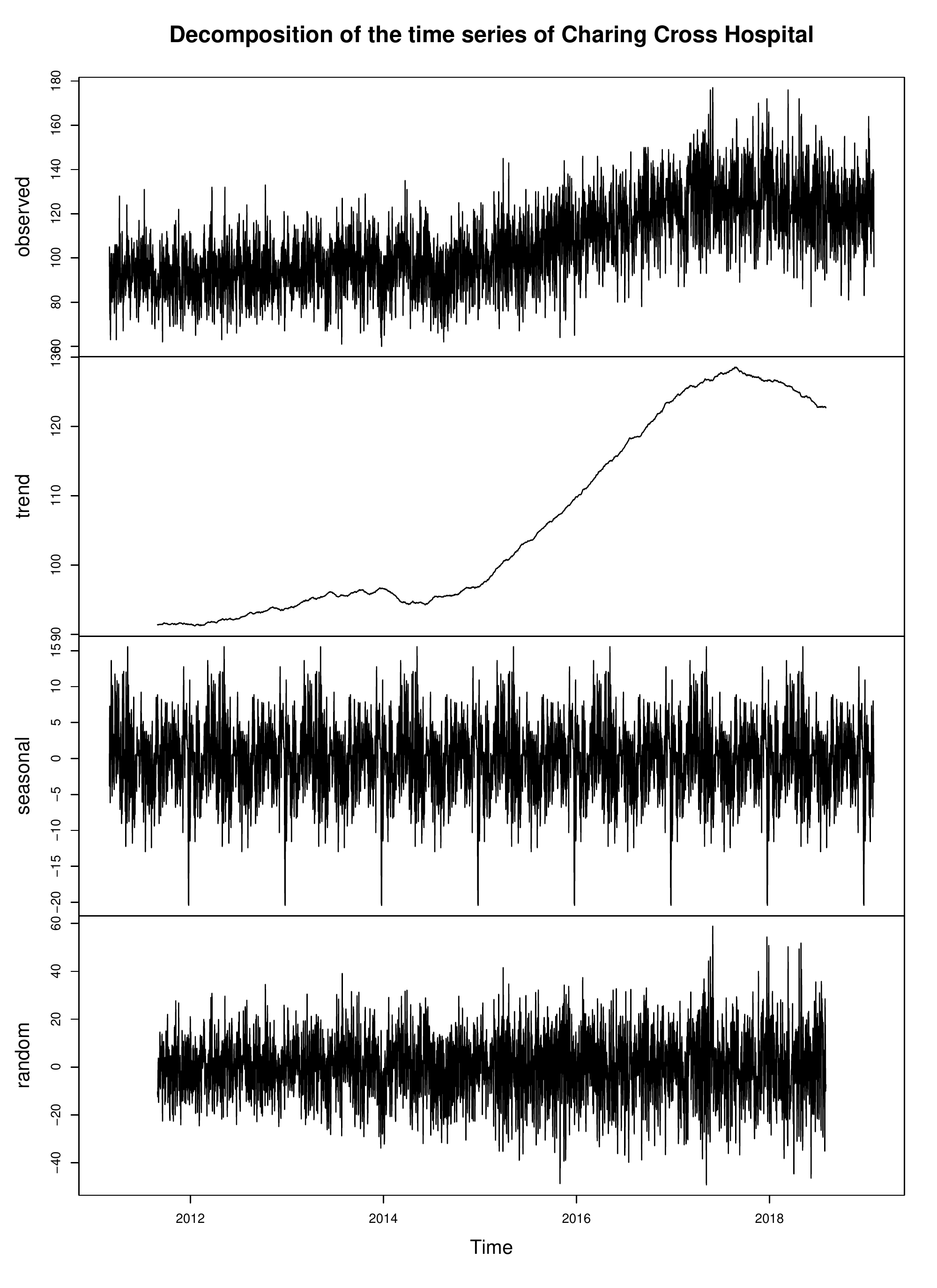}
  \end{subfigure}
  \caption{Time series decomposition's of attendances at both EDs for 2011 to 2019. Left St Mary's Hospital and right Charing Cross Hospital. The first row of plots is the data, the second the trend, the third seasonality and the fourth random residuals.}%
    \label{fig:decomp}%
\end{figure}

\begin{figure}[H]
\centering
\includegraphics[width=0.95\linewidth]{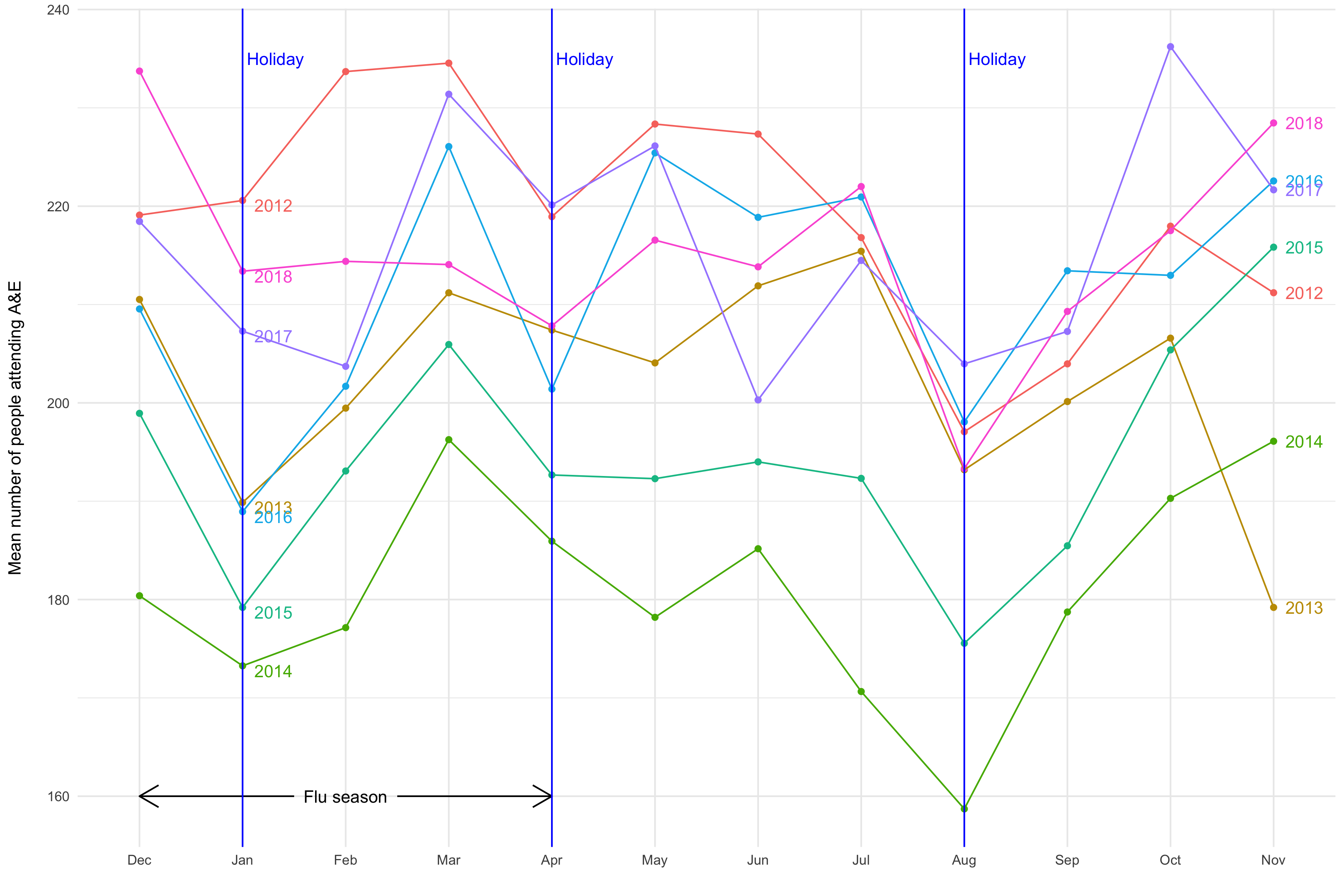}
\caption{Monthly attendance at St Mary's Hospital, 2011-2018.}
\label{fig:montly_SMH}
\end{figure}

A central aim of this paper is not only to predict hospital demand accurately but also attempt to understand the factors driving hospital demand. Daily data is driven by a complex web of exogenous variables, many of which are related to seasonal patterns or trends. Both time series depicted in Figure \ref{fig:timeseries} show a strong underlying trend. In the case of Charing Cross Hospital this trend is clearly upwards while it goes downwards first for St Mary's Hospital due to a change of the hospital's infrastructure, see Figure \ref{fig:decomp}. There is also clear seasonality, the monthly attendance at St Mary's Hospital shows a very clear monthly periodic pattern with troughs in January, April and August and a rise in attendance during the winter months (likely due to increases in acute respiratory infections), see Figure \ref{fig:montly_SMH}. 

\begin{figure}[H]
\centering
\includegraphics[width=0.95\linewidth]{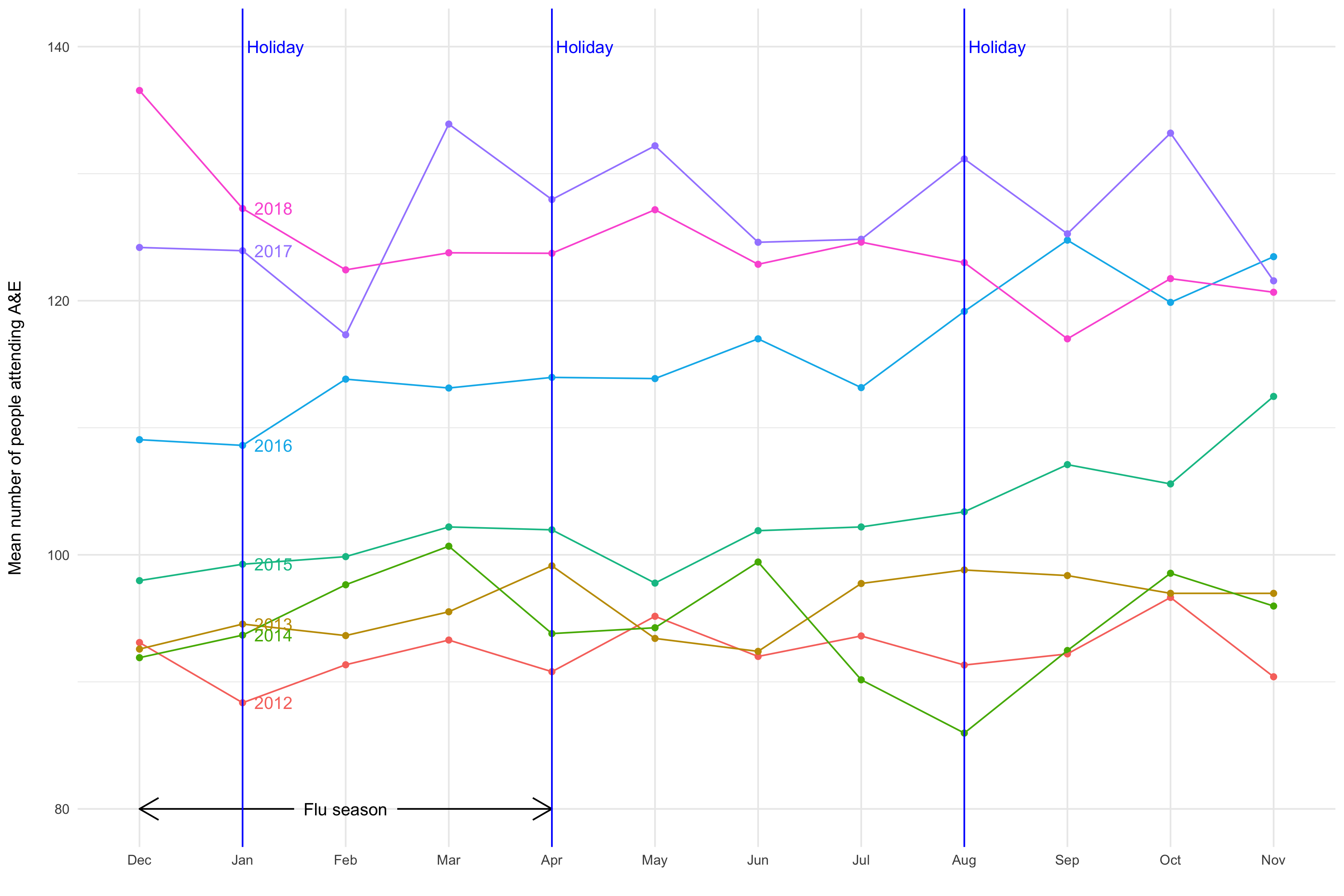}
\caption{Monthly attendance at Charing Cross Hospital, 2011-2018.}
\label{fig:monthly_CCH}
\end{figure}
\begin{figure}
  \begin{subfigure}[b]{0.45\textwidth}
    \includegraphics[width=\textwidth]{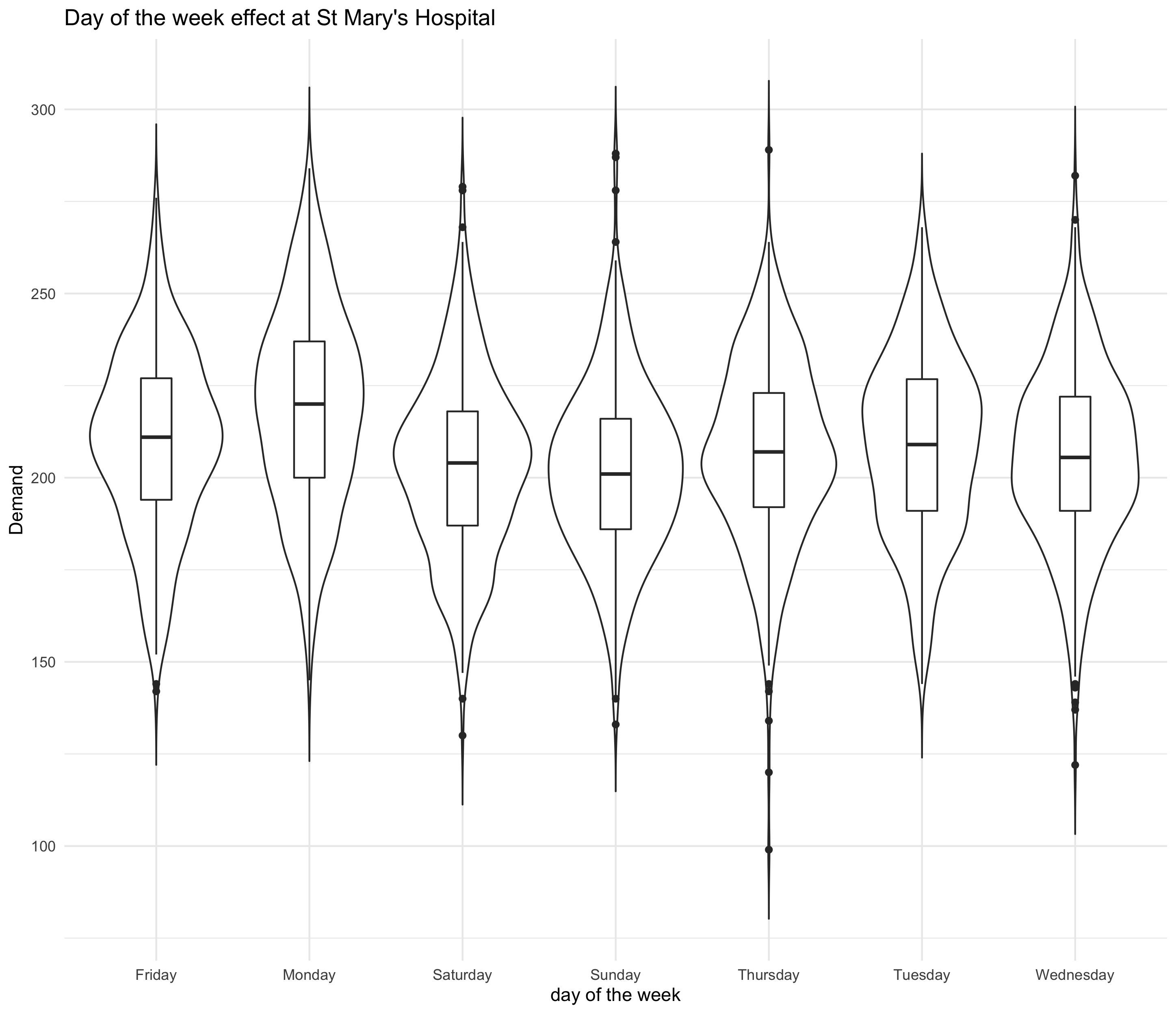}
  \end{subfigure}
  \begin{subfigure}[b]{0.45\textwidth}
    \includegraphics[width=\textwidth]{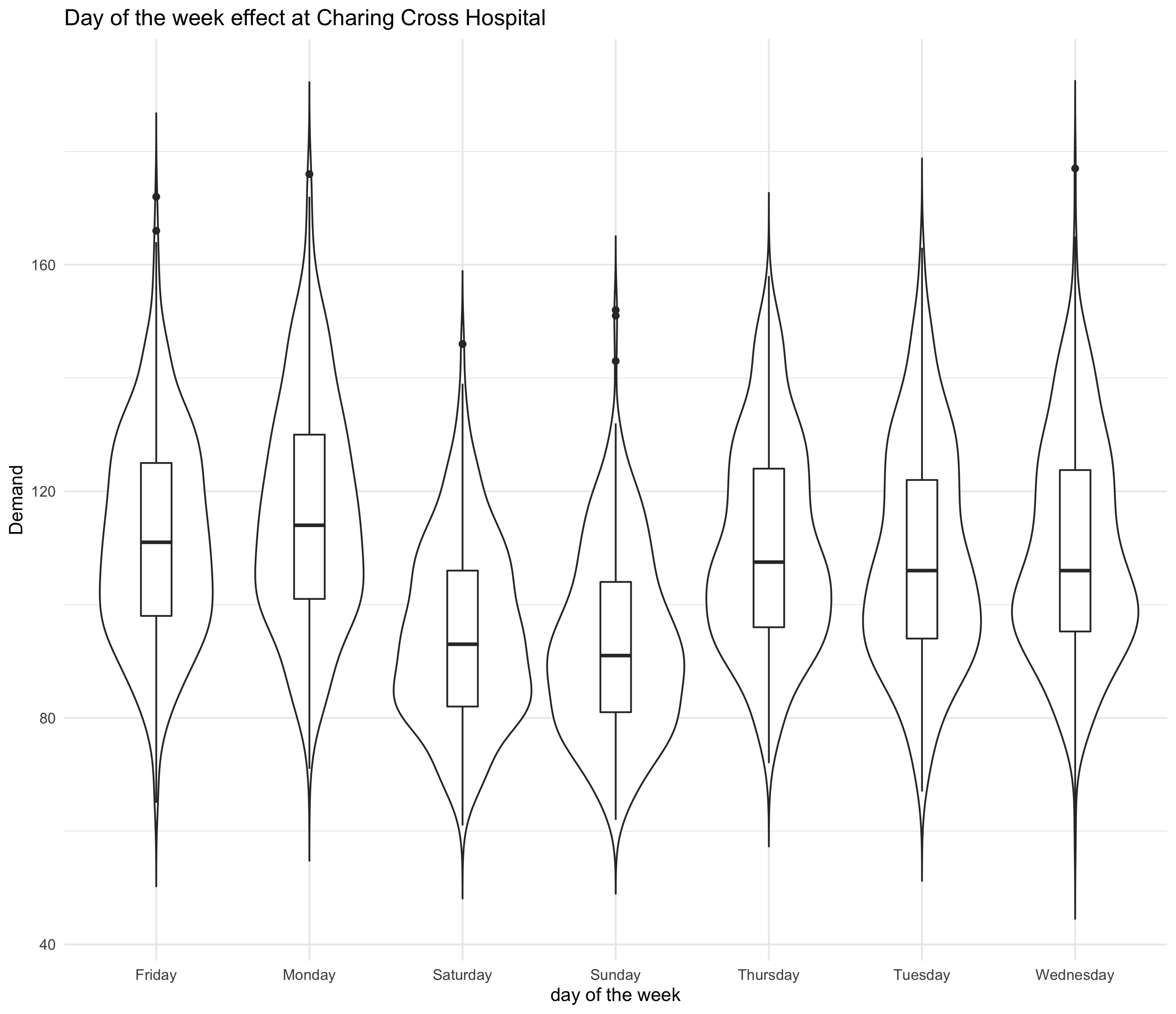}
  \end{subfigure}
  \caption{Day of the week effect for ED attendances at both hospitals 2011-2019.}%
    \label{fig:weekdays}%
\end{figure}

Figure \ref{fig:montly_SMH} also shows indications that bank or school holidays have a strong influence on the number of ED admissions together with the flu season. It also shows that the flu seasons contribution to increased demand runs well into spring. While the monthly attendance at Charing Cross Hospital also shows some periodic behaviour, it is not as strong, see Figure \ref{fig:monthly_CCH}. It is therefore useful to note that dynamics differ even from geographically close hospitals with overlapping catchments. Both series show clear day-of-week patterns, characterised by a strong autocorrelation with respect to their lagged values of order $7$, see Figure \ref{fig:weekdays}. Mondays have the highest volume of attendances at both hospitals while attendance reaches its minimum during weekends. This finding validates and confirms other studies on hospital demand \cite{NHS08-09,NHS17-18}.

\section{Methods and algorithms}
\label{algorithms}
We focus on forecasting demand one, three, and seven days into the future. These particular forecasting intervals are relevant as they allow the hospitals to take action by using short term measures such as the cancellation of elective surgeries or the hiring of additional staff through agencies.

We use two different kinds of algorithms for our predictions: traditional time series and machine learning algorithms \cite{M4}. A discrete time series is a sequence of data points in chronological order divided into regular time intervals. The fundamental assumption behind both algorithms is that data points that are close to each other in time show a similar behaviour and that there is a dependency between data points at the same position of the time interval, e.g. same time of the year or same day of the week. Time series algorithms use both the chronology of the events and the specified interval in order to make inference and split the time series into different linear components such as seasonality, trend and a residual. The residual, which is assumed to contain some correlative structure, is usually modelled using an autoregressive stochastic process or exponential decay where future values are predicted based on past values \cite{M4}. In contrast, machine learning algorithms specify a broad function class (such as trees or smooth curves) with sufficient capacity to learn complex functions. These algorithms learn from data balancing function complexity with predictive accuracy. For both sets of algorithms we create a model (design) matrix containing explanatory variables. For the time series algorithms only lagged demand from previous time points were used. For the machine learning algorithm lagged demand values were used alongside other covariates (see  Table \ref{tbl:ModelMatrix}). As predictors we have chosen past values such as demand on the previous day, last week and the average of the past week as well as indicators for bank holidays and school holidays. Moreover, we use data from some of the surrounding weather stations on precipitation, minimal and maximal temperature as covariates. Finally, we use search engine query data as a covariate, as it has proven to be a very efficient measure for the detection of influenza epidemics \cite{ginsberg2009}. Table \ref{tbl:ModelMatrix} shows a few rows of our model matrix. 

\begin{table}[!htbp] \centering 
  \caption{Excerpt from the model matrix corresponding to St Mary's Hospital. Explanatory covariates for the machine learning algorithms are listed.} 
  \label{tbl:ModelMatrix} 
\begin{tabular}{@{\extracolsep{5pt}} cccccc} 
\\[-1.8ex]\hline 
\hline \\[-1.8ex] 
Date & 2014-04-10 & 2014-04-11 & 2014-04-12 & 2014-04-13 & \dots \\ 
\hline \\[-1.8ex] 
demand & $198$ & $183$ & $172$ & $185$ & \dots\\
month & April &April&April&April&\dots\\
yesterday & $218$ & $198$ & $183$ & $172$ & \dots\\ 
same day last week & $223$ & $189$ & $187$ & $195$& \dots \\ 
average of previous week & $199.6$ & $196$ & $195.1$ & $193$ & \dots\\ 
time &$-0.306$ &$-0.305$&$-0.304$&$-0.303$& \dots\\
bank holiday & FALSE & FALSE & FALSE & FALSE& \dots \\ 
school holiday & FALSE & FALSE & TRUE & TRUE & \dots\\ 
day of week & Thursday & Friday & Saturday & Sunday & \dots\\ 
precipitation on previous day & $0$ & $0$ & $0$ & $0$ & \dots\\ 
max temperature on previous day & $17.5$ & $17.2$ & $15.0$ & $17.4$ & \dots\\ 
min temperature on previous day & $12.0$ & $11.8$ & $10.8$ & $11.3$ & \dots\\ 
flu hits on google on previous day & $7.52$ &$4.88$ &$5.12$&$4.08$&\dots\\ 
Notting Hill carnival & FALSE & FALSE & FALSE & FALSE & \dots\\
Christmas & FALSE & FALSE & FALSE & FALSE & \dots\\
\hline \\[-1.8ex] 
\end{tabular} 
\end{table}  

Below, we summarise the time series and machine learning algorithms that we have considered (for detailed mathematical information we refer the reader to \cite{hyndman2018}:
\subsection{Time series algorithms}
        \begin{enumerate}
            \item ARIMA - AutoRegressive Integrated Moving Average \cite{hyndman2018}\\
            An ARIMA model consists of three parts: The autoregressive component (AR) referring to the fact that the indicator of interest is regressed on its own previous values (i.e.~current values of demand depend on past values of demand), the integrated (I) part representing one or several differencing steps to make the time series stationary and the moving average (MA) component indicating that the regression error is a linear combination of past error terms. 
            \item ETS - Exponential smoothing methods \cite{hyndman2018}
            In general exponential smoothing refers to forecasting methods which also regress on lagged values of the target variable. However, it uses exponentially decreasing weights for past observations. The ETS model we employ uses exponential smoothing for error, trend and seasonality. 
            \item STLM - a seasonal decomposition of time series by LOESS (STL)a seasonal decomposition of time series by LOESS (STL)
            STLM is another type of exponential smoothing model. The time series is decomposed into its seasonal components using LOESS (locally estimated scatterplot smoothing) before exponential smoothing is used to model the error and trend component of the time series. Finally the series is re-seasonalized. 
            \item StructTS - Structural Time Series Model \cite{hyndman2018}
            A StructTS model is formulated directly in terms of unobserved components, such as trends, seasonality and exogenous factors that have a natural interpretation. The StructTS model forms a state space which makes it similar to the ARIMA and ETS models.
            \end{enumerate}
        
\subsection{Machine learning algorithms}\label{sec:MLalgorithms}
\begin{enumerate}
            \item glmnet \cite{glmnet} is a penalised generalized linear model with built-in variable selection.
            The glmnet model is an extension of the generalised linear model in which bias (penalty/weight decay/regulariser) is introduced in the form of a mixture penalty consisting of the parameter $\ell_1$ and $\ell_2$ norms. The magnitude of this penalty is tuned to balance overfitting vs.~underfitting, with the goal of reducing variance by introducing some bias.
            \item ranger  \cite{ranger}
             is a fast implementation of random forests \cite{breiman}. Random forests are a bagged ensemble of decision trees. Random forests use bootstrapped random subsets of the covariates variables to build decision trees based on these subsets. Random forests prevent overfitting by averaging to reduce variance. 
            \item Gradient Boosting Machines (gbm)  \cite{gbm} are a  generalized boosted regression model.
            In contrast to random forests which builds a collection of multiple independent decision trees, the decision trees created by gradient boosting machines depend additively on each other. Each new tree is added to an ensemble by improving the previous trees residual error (the functional negative gradient). Hyperparameters are tuned to balance over and under fitting.
            \item k-nearest neighbours (k-NN) \cite{bishop} is a classic supervised learning algorithm based on the idea that data points that are close to each other in the space of covariates should have  similar predictions. Hence, to make a prediction at a given location in covariate space, an average of the labels of the k nearest neighbours is taken. The disadvantage of this algorithm is that it slows down quickly once the volume of data points increases.
    \end{enumerate}

\subsection{Stacked Regression}
All the above algorithms have strengths and weaknesses. It is therefore challenging to choose a single model for prediction. We therefore create a consensus model by adopting stacked regression, a particularly effective ensemble approach. The idea of stacked regression is to combine the diversity and strengths of multiple algorithms into a single model with a better overall ability to generalise. We choose a linear stacking model subject to convex combination constraints \cite{breiman1996stacked}. Following \cite{breiman1996stacked,wolpert1992stacked} we train a linear stacker on cross validation predictions of the individual time series and machine learning models. This procedure helps us avoid selecting models that overfit to the data.
Stacking models is not only empirically motivated, but has a strong theoretical backing and has been proven to perform perform asymptotically exactly as well (by some loss metric) as the best possible choice for the given data set among the family of weighted combinations of the estimators. Stacking has also been showing to work in a variety of settings \cite{Bhatt2017}.

\section{Validation and evaluation}

Developing an algorithm with the best possible forecast accuracy is the main goal of this study. However, care must be taken to ensure that these forecasts are not simply overfitting to noise in the data but are accurate and can truly forecast to unseen data. A key innovation of this paper is the development of a novel general purpose time series cross validation procedure to ensure that: 
\vspace{-12pt}
\begin{enumerate}[label=(\alph*)]
    \item all algorithms are evaluated fairly and equally,
    \item the same data is used in all algorithms, and
    \item forecast errors are completely blind to the held-out data (i.e.~exactly as if the model was being used in a real forecasting setting).
\end{enumerate}
\vspace{-12pt}
Temporal or time series cross-validation \cite{hyndman2018} is a method to split the data into testing and training sets in order to account for temporal structure in the data. The main idea is that each test set only consists of a forecasting window of one day which lies one, three or seven days in the future while the corresponding training set consists of a number of observations prior to the forecasting window. The origin can either be fixed so that the length of the training window grows by one, three, or seven days for each new test set or it can move forward so that the training window size remains fixed. We employ the latter method.

\begin{figure}[H]
\begin{center}
\includegraphics[width=0.8\linewidth]{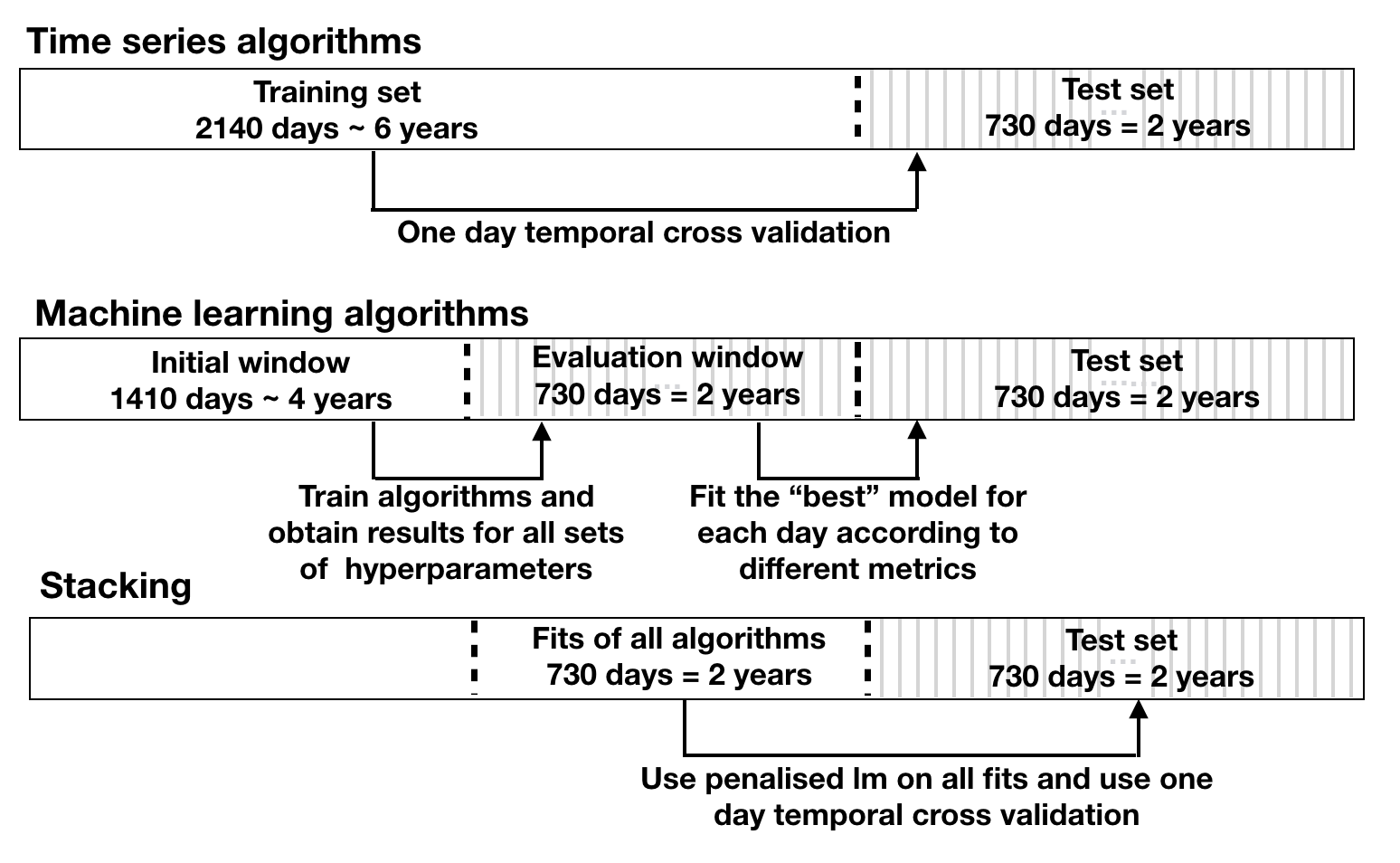}
\end{center}
\caption{Splitting the data into training, (one day) validation and test set.}
\label{fig:datasplit}
\end{figure}

Using temporal cross-validation for time series algorithms only requires splitting the data into a training and test set. Adapting the data so that the last $730$ days ($24$ months) of data are held out for testing, $2140$ days (roughly six years) were available for training which corresponds to a $\sim$  $75$\%/$25$\% split. We then applied temporal cross validation to each day in the test set (see Figure \ref{fig:datasplit}) using a rolling window so that all training sets consist of the same number of days. 

Given the fairness and robustness of our cross validation scheme, we are confident that our results are robust to data shift and not only valid for the data times we have collected.

\subsection{Validation of hyperparameters}
In order to allow for a fair comparison of the machine learning algorithms, the method for temporal cross-validation has to be adapted as most machine learning algorithms require tuning of their hyperparameters to balance over and under fitting. Therefore we also have to split the training data into a training and a validation set in order to choose the best set of hyperparameters which minimizes the error on the validation set.  All machine learning algorithms listed in Section \ref{sec:MLalgorithms} were trained on $4$ years worth of data and their hyperparameters were compared on a validation set consisting of $730$ days. In our analysis we have used two different approaches to split the training set and to choose the hyperparameters. We will call these the batch and the online method.

\subsubsection{The batch method}
\label{sec:batch}
In order to choose the set of hyperparameters which minimizes the error on the validation set, we consider five different approaches. We choose the set of hyperparameters which minimize the error
\vspace{-12pt}
\begin{enumerate}
    \item on the previous day,
    \item on the past $n$ days,
    \item over the whole validation period using an exponential moving average,
    \item averaged over the whole validation period, or
    \item according to caret's built in rules (see \cite{Kuhn2008}).
\end{enumerate}
\vspace{-12pt}
For each of the five cases above, we choose the best set of hyperparameters for each day of the test set and refitted all models on a daily basis. Of course refitting every model for each day of the test set is computationally expensive. Therefore we  develop an online method, described below.

\subsubsection{The online method}
\label{sec:online}
In the batch method described above, all models are refitted daily, which takes significant computational power to run and might not be feasible for a deployed version of our methods in a hospital setting. Thus, we explore whether keeping the parameters fixed for longer periods, which yields significant savings in terms of computation, hurts performance. We refit each model over several testing periods of various sizes which are subsets of the test set of length $730$ days with a rolling origin. Our chosen testing periods are $1$ day, $7$ days, $30$ days, $60$ days, $365$ days and $730$ days long. That means that we validated the best set of hyperparameters for each algorithm for every testing period, and  then rolled forwards. The final error rates are the result of the overall error on all predictions on the whole test set of length $730$.

\begin{figure}[H]
\centering
\includegraphics[width=0.8\linewidth]{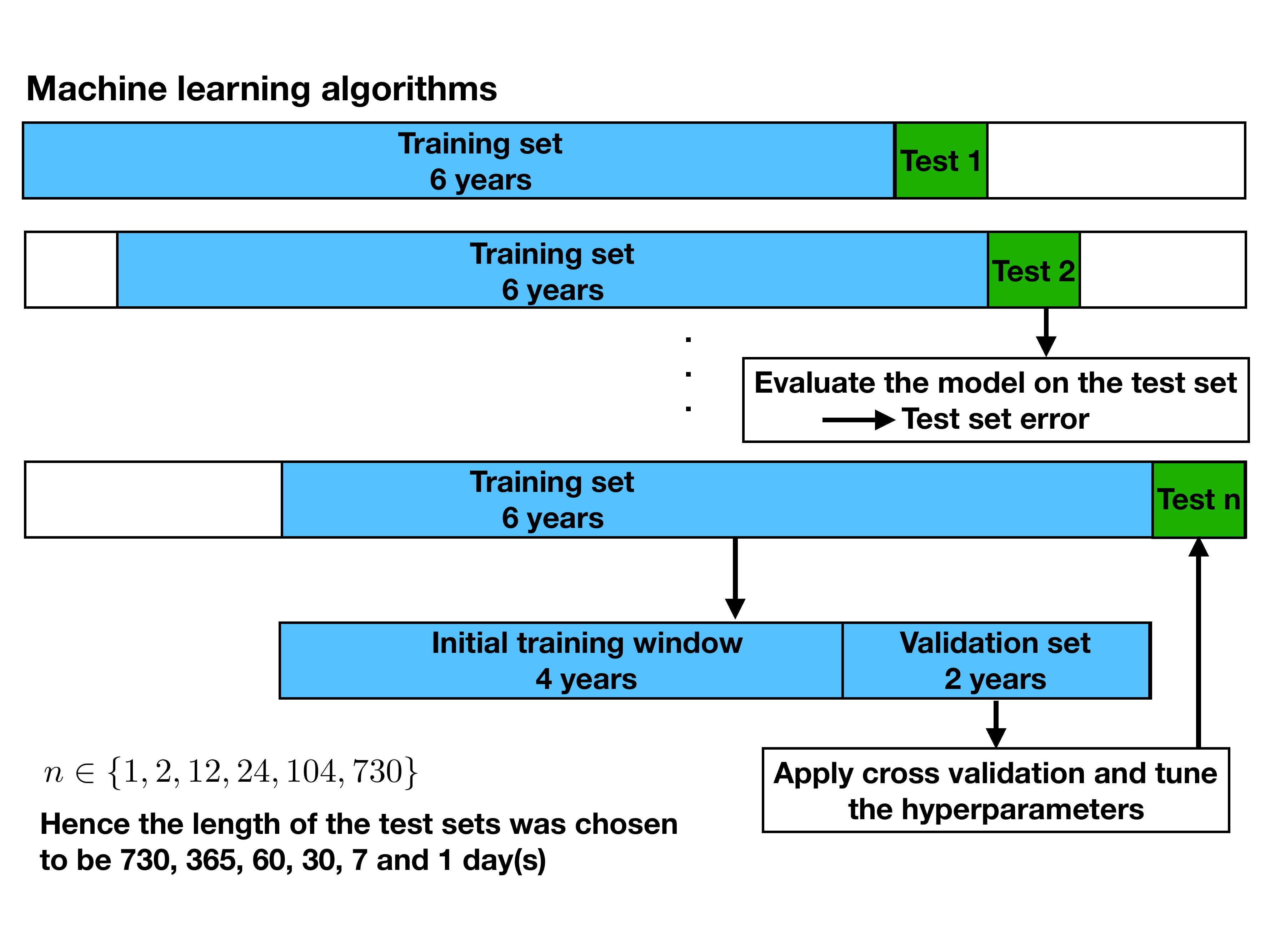}
\label{fig:ML-DataSplit}
\caption{Splitting the data set into training, validation and test set in case of the online method.}
\end{figure}

\section{Results}
We compare the mean absolute error (MAE) and mean absolute percentage error (MAPE) rates for all time series algorithms as well as for the batch and the online method as outlined in Section \ref{sec:batch} and \ref{sec:online}. Finally, we compare our results with the stacked regression.  

\begin{table}[!htbp] \centering 
  \caption{Error rates for all time series algorithms for St Mary's and Charing Cross Hospital.} 
  \label{tbl:TS} 
\begin{tabular}{@{\extracolsep{5pt}} ccccccc}
\\[-1.8ex]\hline \hline \\[-1.8ex] 
&&\multicolumn{2}{c}{\underline{St Mary's Hospital:}}&\multicolumn{2}{c}{\underline{Charing Cross Hospital:}}\\ \\[-1.8ex] 
 algorithm & days & MAE &  MAPE (in\%) & MAE &  MAPE (in \%) \\ 
\hline \\[-1.8ex] 
        &$1$& $15.29$ & $7.2$ & $12.33$ & $10.1$ \\ 
ARIMA   &$3$& $15.68$ & $7.4$ &
        $14.75$ & $12.2$ \\
        &$7$ & $15.58$ & $7.4$ & $13.81$ & $11.4$\\ 
\hline \\[-1.8ex]    
        & $1$ & $14.47$ & $6.9$ & $10.46$ & $8.5$\\
ETS     & $3$ & $16.89$ & $8.0$ &         $17.60$ & $14.6$ \\
        & $7$ & $16.18$ & $7.6$ & $14.55$ & $11.8$\\
\hline \\[-1.8ex] 
        & $1$& $14.54$ &$6.9$ & $10.72$ & $8.7$\\
StructTS& $3$ & $17.35$& $8.2$ &         $17.86$ & $14.8$\\
        & $7$ & $16.54$& $7.8$ & $14.72$ & $12.0$\\
\hline \\[-1.8ex]
        & $1$ & $14.51$ & $6.9$ & $10.59$ & $8.6$\\
STLM    & $3$ & $16.67$  & $7.9$&         $16.83$ & $13.9$    \\ 
        & $7$ & $16.00$ & $7.6$ & $14.28$ & $11.6$\\
\hline \\[-1.8ex] 
\end{tabular} 
\end{table}

\subsection{St Mary's Hospital}
The MAE of the time series algorithms range from $14.46$ to $17$ as shown in Table \ref{tbl:TS}, with an MAPE ranging from from $6.9\%$ to $8.3\%$. Hence, although the time series algorithms are simpler and only based on the time series without using any other predictors, such as weather, bank or school holidays (see Table \ref{tbl:ModelMatrix} for details on all predictors), they already give relatively accurate results. In case of the machine learning algorithms, independent of the type of hyperparameter tuning we use, most of the results for the batch methods range between an MAE of $14.3$  to $15.2$ and an MAPE of $6.8\% - 7.4\%$, as shown in Table \ref{tbl:SMH-batch}. Only the k-nearest neighbour algorithm performs worse despite different ways of tuning. Especially when using the online method, see Table \ref{tbl:OnlineMethod}, the linear models produce the best results for St Mary's hospital with an MAPE of $6.8\%$ in the case of daily retraining of the generalized linear model.

\begin{table}[!htbp] \centering 
  \caption{MAE rates for all types of hyperparameter tuning predicting hospital demand $1$, $3$ or $7$ days in advance in case of the batch method for St Mary's hospital.} 
  \label{tbl:SMH-batch} 
\begin{tabular}{@{\extracolsep{5pt}} ccccccc}
\\[-1.8ex]\hline 
\hline \\[-1.8ex] 
&&\multicolumn{5}{l}{\underline{Choosing the best set of hyperparameters based on:}}\\
 algorithm & days & yesterday &  \begin{tabular}{@{}c@{}}the past\\ $n$ days\end{tabular} 
 &\begin{tabular}{@{}c@{}}  exponential\\ moving average\end{tabular} & \begin{tabular}{@{}c@{}}the average over the\\ whole training set\end{tabular} & caret \\ 
\hline \\[-1.8ex] 
        &$1$& $14.50$ & $14.38$ & $14.48$ & $14.42$ & $14.34$ \\ 
    lm  &$3$& $14.80$ & $14.96$ & $14.86$ & $14.95$ & $14.80$ \\ 
        &$7$ & $15.11$ & $15.17$ & $15.22$ & $15.33$ & $15.13$ \\ 
\hline \\[-1.8ex]    
         & $1$ & $15.97$ & $14.49$ & $15.08$ & $14.46$ & $14.30$ \\
gbm     & $3$ & $15.38$ & $14.63$ & $14.86$ & $14.67$ & $14.53$ 
\\ 
        & $7$ & $15.43$ & $14.79$ & $14.93$ & $14.78$ & $14.86$ 
        \\ 
\hline \\[-1.8ex] 
        & $1$& $14.49$ &$14.34$ &$14.38$ &$14.35$ &$14.48$ \\
glmnet  & $3$ & $14.81$ & $14.79$ & $14.82$ & $14.75$ & $14.77$ \\
        & $7$ & $14.98$ & $15.11$ & $14.94$ & $15.16$ & $15.09$ \\ 
\hline \\[-1.8ex]
        & $1$ &$15.86$ & $15.47$ & $15.74$ & $15.79$ & $15.42$\\
knn     & $3$ & $16.13$ & $15.78$ & $15.92$ & $16.02$ & $15.57$ \\ 
        & $7$ & $16.70$ & $16.05$ & $16.45$ & $16.52$ & $15.89$ \\ 
\hline \\[-1.8ex]       
         & $1$ & $14.62$ & $14.46$ & $14.54$ & $14.79$ & $14.45$ \\
rf      &$3$& $14.66$ & $14.55$ & $14.54$ & $15.15$ & $14.53$ 
\\ 
        & $7$& $14.93$ &$14.61$&$14.62$&$15.53$&$14.67$
        \\
\hline \\[-1.8ex] 
\end{tabular} 
\end{table} 

\begin{table}[!htbp] \centering 
  \caption{MAE error rates for all types of hyperparameter tuning predicting hospital demand $1$, $3$ or $7$ days in advance in case of the batch method for Charing Cross hospital.}
  \label{tbl:CCH-batch} 
\begin{tabular}{@{\extracolsep{5pt}} ccccccc}
\\[-1.8ex]\hline 
\hline \\[-1.8ex] 
&&\multicolumn{5}{l}{\underline{Choosing the best set of hyperparameters based on:}}\\
 algorithm & days & yesterday &  \begin{tabular}{@{}c@{}}the past\\ $n$ days\end{tabular} 
 &\begin{tabular}{@{}c@{}} an exponential\\ moving average\end{tabular} & \begin{tabular}{@{}c@{}}the average over the\\ whole training set\end{tabular} & caret \\ 
\hline \\[-1.8ex] 
        &$1$& $10.66$ & $10.60$ & $10.67$ & $10.58$ & $10.51$ \\ 
    lm  &$3$ & $10.75$ & $10.72$ & $10.84$ & $10.69$ &           $10.69$ \\ 
        &$7$& $10.78$ & $10.90$ & $10.83$ & $10.87$ & $10.77$ \\ 
\hline \\[-1.8ex]    
        & $1$ & $12.50$ & $10.68$ &$11.12$& $10.68$& $10.89$ \\ 
gbm     & $3$ & $12.49$  & $10.58$  & $10.92$  &$10.50$  & $10.78$  \\ 
        & $7$ & $12.11$ & $10.73$ & $11.03$ & $10.57$ & $10.72$  \\ 
\hline \\[-1.8ex] 
        & $1$& $10.72$ & $10.51$ & $10.59$ & $10.51$ & $10.51$ \\
glmnet  & $3$ & $10.85$ & $10.69$ & $10.74$ & $10.69$ & $10.69$ \\
        & $7$ & $10.86$ & $10.76$ & $10.84$ & $10.76$ & $10.76$ \\ 
\hline \\[-1.8ex]
        & $1$  & $12.81$ & $12.57$ & $12.62$ & $12.77$ & $12.36$\\
knn     & $3$ & $12.85$ & $12.58$ & $12.78$ & $12.87$ & $12.51$ \\ 
        & $7$ & $12.65$ & $12.45$ & $12.41$ & $12.47$ & $12.33$ \\ 
\hline \\[-1.8ex]       
        & $1$& $11.09$ & $10.84$ & $11.04$ & $11.55$ & $10.89$ 
        \\ 
rf      &$3$& $10.98$ & $10.80$ & $10.83$ & $11.85$ & $10.78$ 
\\ 
        & $7$& $10.92$ & $10.78$ & $10.79$ & $11.99$ & $10.72$
        \\
\hline \\[-1.8ex] 
\end{tabular} 
\end{table}

\begin{table}[h]
\begin{subtable}[h]{0.45\textwidth}
\begin{tabular}{@{\extracolsep{5pt}} cccc}
\\[-1.8ex]\hline 
\hline \\[-1.8ex] 
 algorithm & period  & MAE & MAPE \\
\hline \\[-1.8ex] 
lm & $1$ & $14.27$ & $6.7$ \\
glmnet & $1$ & $14.31$ & $6.8$ \\ 
gbm & $730$ & $14.33$ & $6.8$ \\ 
lm & $7$ & $14.33$ & $6.8$ \\
lm &$365$ & $14.34$ & $6.8$ \\
glmnet & $7$ & $14.37$ & $6.8$ \\ 
glmnet & $365$ & $14.38$ & $6.8$ \\
gbm & $7$ & $14.40$ & $6.8$ \\
lm & $60$ & $14.47$ & $6.8$ \\ 
glmnet & $60$ & $14.47$ & $6.8$ \\
glmnet & $1$ & $14.49$ & $6.9$ \\
lm & $30$ & $14.50$ & $6.8$ \\ 
glmnet & $30$ & $14.50$ & $6.8$ \\
gbm & $30$ & $14.52$ & $6.9$ \\
gbm & $60$ & $14.52$ & $6.9$ \\
gbm & $365$ & $14.53$ & $6.9$ \\
rf & $1$ & $14.55$ & $6.9$ \\
glmnet &$730$ & $14.58$ & $6.8$ \\
lm & $730$ & $14.60$ & $6.9$ \\
rf & $7$ & $14.66$ & $6.9$ \\ 
rf & $730$ & $14.73$ & $6.9$ \\ 
rf & $365$  & $14.76$ & $7.0$ \\ 
rf & $60$ & $15.02$ & $7.1$ \\ 
rf & $30$ & $15.08$ & $7.1$ \\
knn & $1$ & $15.52$ & $7.3$ \\ 
knn & $7$ & $15.53$ & $7.3$ \\ 
knn & $730$ & $15.61$ & $7.4$ \\ 
knn & $365$ & $15.62$ & $7.4$ \\ 
knn & $30$ & $15.75$ & $7.4$ \\ 
knn & $60$ & $15.93$ & $7.5$ 
\end{tabular}
\caption{St Mary's Hospital}
        \label{tab:week2}
\end{subtable}  
\begin{subtable}[h]{0.45\textwidth}
\begin{tabular}{@{\extracolsep{5pt}} cccc}
\\[-1.8ex]\hline 
\hline \\[-1.8ex] 
algorithm & period &MAE & MAPE \\ 
\hline \\[-1.8ex] 
lm & $730$ &  $10.59$ & $8.6$ \\
lm & $1$ &  $10.59$ & $8.6$ \\
glmnet  & $1$  & $10.60$ & $8.6$ \\
lm & $7$ & $10.62$& $8.6$ \\ 
gbm & $1$ & $10.63$& $8.6$ \\ 
glmnet  & $7$  & $10.64$& $8.6$ \\ 
glmnet & $730$  & $10.64$  & $8.5$ \\
gbm & $7$ & $10.71$& $8.7$ \\ 
gbm & $730$ & $10.78$& $8.7$ \\ 
lm  & $365$ & $10.80$ & $8.7$ \\ 
glmnet & $365$ & $10.82$ & $8.7$ \\
lm & $30$ & $10.84$ & $8.8$ \\ 
glmnet & $30$ & $10.84$ & $8.8$ \\
lm & $60$ & $10.86$ & $8.8$ \\ 
glmnet & $60$ & $10.87$ & $8.8$ \\
rf & $1$ & $10.93$ & $8.9$ \\ 
gbm  & $365$ & $10.96$& $8.9$ \\
gbm  & $30$ & $11.00$& $8.9$ \\
rf & $7$ & $11.08$ & $9.0$ \\ 
gbm  & $60$ & $11.15$& $9.0$ \\
rf  & $60$ & $11.21$ & $9.1$ \\
rf  & $30$ & $11.21$ & $9.1$ \\
rf  & $365$ & $11.49$ & $9.2$ \\
rf  & $730$ & $11.51$ & $9.0$ \\
knn  & $7$ & $12.55$ & $10.1$ \\ 
knn  & $1$ & $12.60$ & $10.1$ \\
knn  & $30$ & $12.75$ & $10.2$ \\ 
knn & $60$ & $12.78$ & $10.2$ \\
knn  & $730$ & $12.81$ & $10.1$ \\
knn  & $365$ & $12.81$ & $10.1$ 
    \end{tabular}

\caption{Charing Cross Hospital}
  \label{tab:cch}
\end{subtable}
\caption{Results for the online method for both hospitals.}
\label{tbl:OnlineMethod}
\end{table}

\subsection{Charing Cross Hospital}
The results for Charing Cross Hospital are similar although the error rates are slightly increased. As shown in Table \ref{tbl:TS}, most time series algorithms yield MAE error rates between 10.5\% and 14.5\% while the corresponding MAPE error rates range from 8.5\% to 12\%. The best performance reached using the batch method is also around 10.5\% although the overall performance is a little bit better, see Table \ref{tbl:CCH-batch}. In case of the online method, the MAPE error rates are as low as $8.6\%$ for gradient boosting machine, the generalized linear model or a simple linear model retrained on a daily basis, as shown in Table \ref{tbl:OnlineMethod}.

\subsection{Stacked Regression}
In order to make use of the strengths of all algorithms, we applied a generalised linear model as well as a penalized regression to create an effective ensemble approach, see Figure \ref{fig:datasplit} for details. The best performance was achieved using penalized regression with an MAE error rate of $14.51$ for St Mary's Hospital and $10.60$ for Charing Cross Hospital.

\subsection{Interpretation of results}
In our analysis we consider a variety of predictors ranging from past values of ED attendance, to the weather forecast and to school and bank holidays. The question left to answer is therefore which of the predictors are actually important for making accurate predictions? Could some of them actually be redundant?

\begin{figure*}
        \centering
        \begin{subfigure}[b]{0.475\textwidth}
            \centering
            \includegraphics[width=\textwidth]{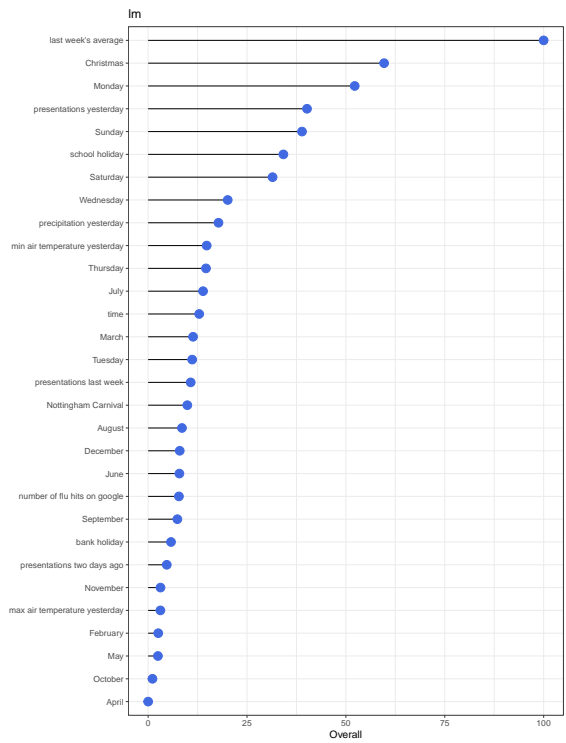}
            \caption[lm]%
            {{\small lm}}    
            \label{fig:mean and std of net14}
        \end{subfigure}
        \hfill
        \begin{subfigure}[b]{0.475\textwidth}  
            \centering 
            \includegraphics[width=\textwidth]{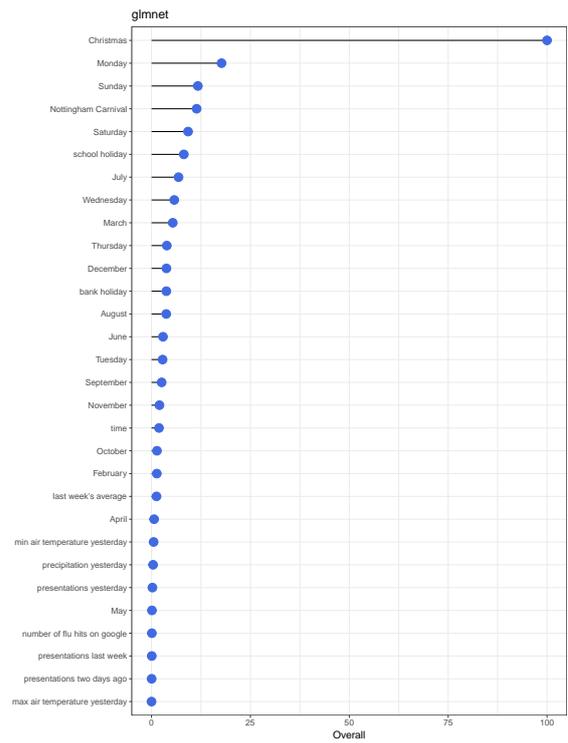}
            \caption[glmnet]%
            {{\small glmnet}}    
            \label{fig:mean and std of net24}
        \end{subfigure}
        \vskip\baselineskip
        \begin{subfigure}[b]{0.475\textwidth}   
            \centering 
            \includegraphics[width=\textwidth]{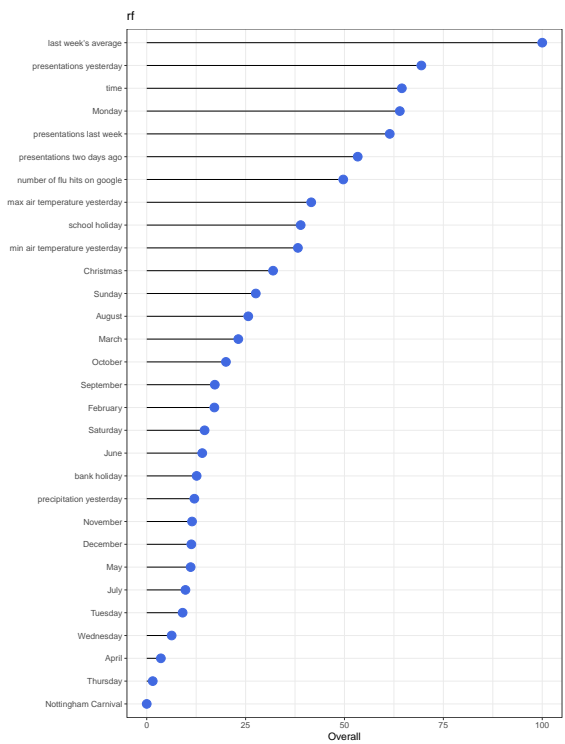}
            \caption[]%
            {{\small gbm}}    
            \label{fig:mean and std of net34}
        \end{subfigure}
        \quad
        \begin{subfigure}[b]{0.475\textwidth}   
            \centering 
            \includegraphics[width=\textwidth]{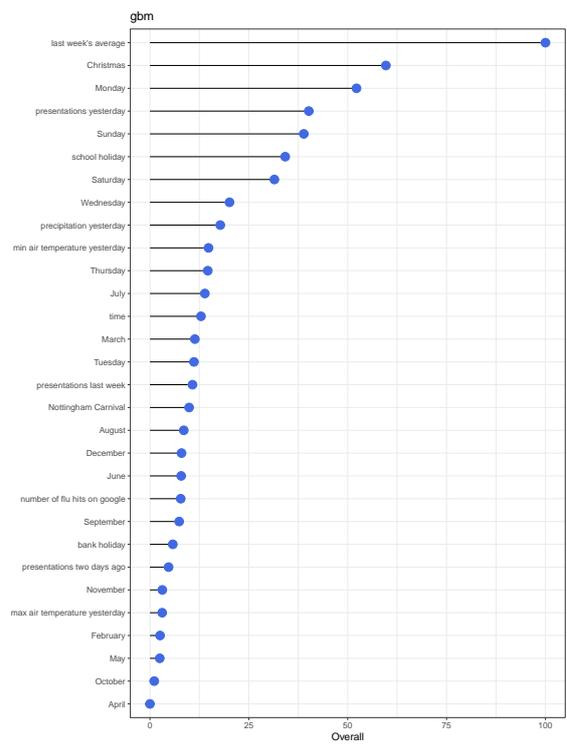}
            \caption[]%
            {{\small rf }}    
            \label{fig:mean and std of net44}
        \end{subfigure}
        \caption[ ]
        {\small Variable importance for the machine learning algorithms.} 
        \label{fig:varImp}
    \end{figure*}

In machine learning, variable importance can be defined as the dependence between input and output variables and  computed by permuting the values of a given predictor  and calculating error on a held out set. This measure has drawbacks in the case of multicollinearity, which could suppress the importance of certain variables in some models.  As shown in Figure ~\ref{fig:varImp}, the most important variable predicting demand is the average demand the week before - except for glmnet - followed by specific days and months. The importance of these variables varies considerably between algorithms but largely is consistent with one another. To our knowledge, this is the first attempt to quantify the importance of variables in predicting demand. We believe that these importance percentages can be utilised as heuristics to help staff improve their prediction of demand in the absence of statistical analysis.

\section{Conclusion}
The results of our analysis highlight interesting statistical properties: simple linear methods like generalized linear models are often better or at least as good as ensemble learning methods like the gradient boosting or random forest algorithm. However, though sophisticated machine learning methods are not necessarily better than linear models, they improve the diversity of model predictions so that stacked predictions are more robust and accurate than any single model including the best performing one. This largely confirms a 'wisdom of the crowd' rule for ensemble learning.

The online method we have created has the ability to provide accurate results with a very quick turnaround time: an average model only takes a couple of minutes to train and forecast. Running it over longer periods of time without retraining of the model is much faster and at the same time not significantly worse than tuning the hyperparameters on a daily basis.

The framework, analysis and methodology proposed in this paper are highly relevant from an operational viewpoint. To the authors knowledge, the majority of EDs in the UK do try to account for demand, but do so using ad hoc heuristics. Our approach provides some scientifically backed information to improve these heuristics, but more importantly provides a framework that is quick and easy to implement. Estimates of $1$, $3$ and $7$ day demand forecasts can be created at any time and updated easily allowing statistically backed estimates to be used to inform hospital policy and practice. Our hope is this paper will fill a knowledge gap and increase hospitals uptake in the use of these methods. Given epidemics and disease outbreaks that can strain health systems, our approach can provide added precision to help EDs operate as efficiently as possible. The challenge will then be for teams in hospitals to implement more insight-driven ways of working, more flexible approaches to rostering staff and innovative ways of communicating with their local communities.

In an era of precision medicine, future work will undoubtedly focus on granular data sets including individual level patient data with diagnoses and testing information. This will not only help to improve performance of the predictions but also help to understand dependencies between seasonality, events and reasons for ED presentations. This may come at the cost of longer running times but only an in depth analysis of patient's pathways, the rate of admission for different diagnoses and their corresponding bed occupancy will help improve hospital planning substantially.

\section{Acknowledgements}
We acknowledge the NIHR Imperial BRC, joint Centre funding from the UK Medical Research Council and Department for International Development. Grant reference: MR/R015600/1. The authors would like to thank Claire Hook (Director of Operational Performance) for helping understand the operational impact of this project. GC is supported in part by an NIHR research  professorship. CC is funded by a National Institute for Health Research (NIHR) Career Development Fellowship (NIHR-CDF-2016-09-015) and NIHR North West London Applied Research Collaborative funding (NIHR200180). The views expressed are those of the author(s) and not necessarily those of the NIHR or the Department of Health and Social Care.

\section{Competing interests}
The authors declare that they have no competing interests.

\section{Appendix}
For weather data we used minimum and maximum daily temperatures as reported by the Met Office \cite{METrainfall, METtemperature} from the following weather station:
\begin{itemize}
    \item ID $697$ St. James's Park (postcode SW1A 1), straight line distance: $2.3$ miles/$3.7$ km to St Mary's hospital and $4.23$ miles/$6.81$ km to Charing Cross Hospital.
\end{itemize}
Data on the daily precipitation in London is publicly available \cite{rainfall}. 

The number of Google searches for ``flu'' on the google search engine is also publicly available and can be downloaded from Google trends \cite{GoogleTrends}. Google provides data on the relative search volume and assigns a score between 1 and 100 to every time unit within the time frame specified by the user. However, the length of this time unit is specified by Google, so the data within the time frame may only show the weekly or monthly searches for "flu". Therefore we have to adjust the data by taking both the daily data, downloaded within a time frame of six months, and monthly data, downloaded over the past few years, into account \cite{GoogleTrends-adjust}.

\bibliographystyle{plainurl}
\bibliography{bibliography.bib}

\end{document}